\magnification=1200
\vsize=8.5truein
\hsize=6truein
\baselineskip=20pt
\noindent{\bf SOLUBILITY AND MIXING IN FLUIDS}

\noindent{\it Robert Ho{\l}yst,
Institute of Physical Chemistry, Polish Academy of Sciences,
Poland}
\item{} {\bf Introduction}\dotfill 
\item{\bf 1.} {\bf Thermodynamics}\dotfill 
\item{1.1} {Colligative Properties}
\dotfill  
\item{1.2} {Binary Mixtures: Liquid-Vapor Coexistence}\dotfill 
\item{1.3} {Partially  Miscible Liquids}\dotfill 
\item{\bf 2.} {\bf Statistical Mechanical Theories of Mixtures}\dotfill 
\item{2.1} {Homogeneous Systems and Ideal Mixtures}
\dotfill  
\item{2.2} {Homogeneous Systems  and Excess Mixing Functions}\dotfill 
\item{2.3} {Conformal Fluid Theories and One Fluid Approximation}\dotfill 
\item{\bf 3.} {\bf Interaction Potential and Mixing }\dotfill 
\item{3.1} {Van der Waals Interactions}
\dotfill  
\item{3.2} {Steric Interactions}\dotfill 
\item{3.3} {Hydrogen Bonding}\dotfill 
\item{\bf 4.} {\bf Polymer Blends and Polymers in Solutions}\dotfill 
\item{4.1} {Polymer Blends }\dotfill 
\item{4.2} {Polymers in Solutions}
\dotfill  
\item{\bf 5.} {\bf Ordering and Demixing}\dotfill 
\item{5.1} {Mixtures of Liquid Crystals}\dotfill 
\item{5.2} {Diblock Copolymers}\dotfill
\item{5.3} {Amphiphilic Systems}\dotfill
\item{\bf 6.}  {\bf Kinetics of Demixing}\dotfill
\item{} {\bf Glossary}
\item{} {\bf Works Cited}
\vfill\eject
\noindent {\bf INTRODUCTION}

The standard 
definition (Mc-Graw Hill Dictionary (1984)) of solubility is as follows:
It is the ability of the substance to form a solution with another substance.
The solution is a single homogeneous liquid, solid or gas phase that is a 
mixture in which the components
(liquid, gas, solid or  the combination thereof) 
are uniformly distributed throughout the mixture. Finally, miscibility
denotes the tendency or capacity of two or more 
liquids to form a uniform blend,
that is to dissolve in each other. For historical reasons
the substance
less abundant in the mixture is called a solute, while the
most abundant one a solvent. 
Here we shall concentrate on nonelectrolytes i.e. solutions in which
none of the components is in the form of free ions.

Traditionally, solubility and mixing belong to the realm of chemistry
and material sciences and many standard textbooks on physical chemistry
(Nernst (1904),
Atkins (1993)) or chemistry (Grant and Higuchi (1990), James (1986))
treat this problem, usually within the scope  of thermodynamics. Here
apart from thermodynamics we shall discuss 
other issues such as: the statistical mechanical theories of mixtures,
relation between intermolecular interactions and demixing,
coupling between ordering and demixing, and 
kinetics of demixing which includes
spinodal decomposition in binary liquid mixtures.
The following special examples will be discussed: polymer blends, 
diblock copolymers,
liquid crystals, and ternary mixtures including surfactants (amphiphiles).
Within the scope of physics of solubility and mixing
one can study
such diverse phenomena as: mixing of two simple liquids,
collapse of the polymer chain in the solvent, flocculation of colloidal particle
upon the addition of the polymer chains, mixing of two polymer 
components in a liquid state,
ordering of copolymers and ternary mixtures of oil, water and surfactant
or formation of micelles in aqueous solutions.
In reality, very rarely we have completely pure substances, thus the properties
of mixtures and the phenomenon of mixing and demixing are of prime importance
for science and technology. 

An old alchemist maxim, ``similia similibus solvuntur''
(``like dissolves like''), is  the oldest rule of solubility.
This rule can be a very good guide in the study of mixing and solubility
providing one can precisely define what in a given case
is the degree of likeness.
For example two simple liquids of  low molecular mass  can easily mix,
say, at room temperature,
while after polimerization they demix  well above  the room temperature.

\noindent{\bf 1. THERMODYNAMICS}

Solubility also means the maximum 
amount of the solute that can be solubilized in the
solvent at the given thermodynamic conditions. 
Within thermodynamics one formulates simple rules 
for solubility of gases and solids in liquids
(Hildebrand and Scott (1950)): 

\item{1.} The solubility of a gas is proportional to its partial
vapor pressure (Henry law). 
\item{2.} The gas with the  higher critical temperature, and boiling point,
is more soluble than one with a lower critical temperature.
\item{3.} The solubility of a gas diminishes with increasing temperature.
\item{4.} The solubility  of a solid increases with increasing temperature.
\item{5.} A solid having a higher melting point is less soluble at a given
temperature than the one having a lower melting point, providing 
that the 
enthalpies of melting are comparable.
\item{6.} A solid with large enthalpy of melting is less soluble
at a given temperature
than the one with small enthalpy of melting, providing
that the melting 
temperatures are comparable.
\item{7.} If the rule 1. is satisfied for a solute in two immiscible solvents
in contact, then the ratio of its concentration in these solvents 
is, for a given
pressure and temperature, a constant (Nerst distribution law).

These rules are not always satisified, 
sometimes only for very dilute solutions.
Nonetheless, in many cases they provide very valuable informations 
about solubilities of different substances in different solvents
based on few thermodynamic properties of pure substances.
All these rules can be expressed in simple mathematical forms 
(Atkins (1993), Hildebrant and Scott (1950)) e.g. 
the rules 4.,5.,6. can be expressed as follows:
$$\ln x={{-\Delta H^s_{solute}}\over R}\left({1\over T}-{1\over {T_m}}\right),\eqno(1)$$
where $x$ is the mole fraction of the solute in the saturated solution
(solution in equilibrium with the solute solid), 
$\Delta H^s_{solute}$ 
is the enthalpy of melting for the solute and $T_m$ is its melting 
temperature. $R$ is the gas constant and $T$ the temperature.

\noindent{\bf 1.1 Colligative properties}

The 
elevation of the boiling point, the depression of
the freezing point in a solution and the osmosis 
are properties which in the first approximation 
do not
depend on the specific nature of the solute, but only on its
amount in the solvent. Such properties are called colligative properties.
For a dilute solution  we find that the addition of the solute into the
pure solvent rises its boiling temperature by:
$$\Delta T=\left({{RT_b^2}\over{\Delta H^v_{solvent}}}\right)x,\eqno(2)$$
where $T_b$ is the boiling temperature of the pure solvent and 
$\Delta H^v_{solvent}$ is the enthalpy of vaporization for the pure solvent.
Similarly the addition of small amount of solute into the solvent
lowers its freezing temperature by
$$\Delta T=-\left({{RT_m^2}\over{\Delta H^s_{solvent}}}\right)x,\eqno(3)$$
where $T_m$ is the freezing temperature of the pure solvent.

The phenomenon of osmosis takes place when the solution is separated
from the pure solvent by a semipermeable membrane, which allows the
flow of a solvent from the pure phase to the solution. The pressure 
that has to be
applied to stop this flow is called the osmotic pressure, $\Pi$. For a
dilute solution one gets van't Hoff equation relating 
this pressure, the volume
of the solution, $V$, and the number of moles of the solute in the solution
$n$,:
$$\Pi V=nRT.\eqno(4)$$
Since $n/V$ is given by the total mass of the solute divided by the molar mass
$M$ and the volume, combination of the measurements of the osmotic pressure
and van't Hoff formula allows to determine the molar mass of the solute.
This method is particularly useful for determining the mass of macromolecules,
however, in  this case it is often necessary to include in Eq(4) the 
next term in
the virial expansion of the osmotic pressure. 

\noindent{\bf 1.2 Binary mixtures: liquid-vapor coexistence}

In the ideal mixture the components 
mix in all proportions without any 
change of volume or enthalpy. The partial 
vapor pressure in the ideal binary mixture
of A (or B) component coexisting with
the ideal liquid mixture 
is given by $P_A=x_A(g)P$ ($P_B=x_B(g)P$).
The total pressure $P$ is  related to the (Raoult law)
vapor pressures above pure A ($P_A^*$) and pure B ($P_B^*$)
liquids:
$$P=P_A+P_B=x_A(l)P_A^*+x_B(l)P_B^*,\eqno(5)$$
These two equations define the vapor ($x_A(g)=1-x_B(g)$)
and the liquid ($x_A(l)=1-x_B(l)$) composition 
at coexistence. In Fig.1 
the composition pressure diagram is shown; the region between the two
curves (given by the above equations) 
is the two phase region, where the relative amounts of 
vapor and liquid are given by the lever rule (Atkins (1993)).

This type of diagram (now temperature and composition, Fig. 2)
is the basis of fractional distillation. We start with the A,B mixture 
at point 1 on the diagram and 
evaporate it (1-2 dashed line). Next we condense it again
until we reach point 3. The resulting liquid mixture is now 
richer in component B. Repeating 
these steps (evaporate it along the 3-4  dashed line etc)
we can 
obtain almost pure B phase. 
Analogous process is used to purify 
solids and is called the zone refining.
In many mixtures the process of distillation is stopped at a
certain point where the 
composition of the vapor is the same as the composition 
of the coexisiting liquid (Fig.3,4). This point is called the
azeotrope. Since at this point the 
liquid boils without changing its composition,
thus the fractional distillation proceeds only until 
positive azeotrope is met
(Fig.4). On the example shown in Fig.3 we reach a 
negative azeotrope by straight
distillation, i.e. by continuously removing the vapor 
from the vessel of the boiling liquid mixture.
For example, the mixture of ethanol and water has the azeotrope at 
$T=78^\circ$C
and at 4\% of water (by mass); water and nitric acid
has the azeotrope at $T=122.4^\circ$C
and $60$\% of HNO$_3$ (mole fraction) (at 1 atm.);
hydrogen chloride and water has the azeotrope at
$T=108^\circ$C and $20$\% (by mass) of HCl (at 1 atm). 
Of course, the location of the azeotrope changes with pressure.
The following rules can help us in deciding when we can expect an azeotrope
in the liquid mixture (Rowlinson and Swinton (1982)):
\item{1.} The closer the vapor pressures of the pure components of the
mixture the more likely is azeotropy. It is inevitable at any temperature
at which they are the same. It also means that when the vapor pressures are
very close then even small departures from ideality of the mixture can produce
an azeotrope.
\item{2.} The closer 
the vapor pressures of pure components the more rapidly does 
the azeotropic composition change with temperature.
\item{3.}  An increase of temperature and of vapor pressure in a positive 
azeotrope (maximum on the p-x diagram  and minimum on the T-x diagram
(Fig.4)) increases the mole fraction of the component whose vapor pressure
increases the more rapidly with temperature. The converse holds for a
negative azeotrope (Fig.3).

These rules are not without exceptions, but 
usually can be a very useful guide. 

\noindent{\bf 1.3 Partially miscible liquids}

Mixtures are often far from being ideal and consequently at suitable
conditions we can expect demixing in the liquid state.
In some  systems demixing 
can take place as we lower the temperature (Fig.5) 
or as we rise the temperature (Fig.6). 
More complicated cases are shown in
Figs.7-10. (Atkins 1993, Landau and Lifshitz (1980)). 
The maximum (minimum) of the curve in Fig.5 (Fig.6) 
is called the upper (lower) critical (or consolute)
point. Gases usually mix very well at normal pressures, however, at
very high pressures they can demix;
then sometimes the phrase gas-gas immiscibility is used
(Rowlinson and Swinton (1982)). The densities of these fluid  phases
at high pressures are comparable to the density of the liquid phases
at `ordinary' low pressures critical points.

Fig.10 is very similar to the 
liquid-solid phase diagram for the binary mixture (instead of vapor we
have liquid
and instead of liquid we have solid). 
The lowest temperature of the liquid mixture at freezing in this case
is obtained for the eutectic composition (point E on the diagram).

For ternary mixtures the phase diagram is often represented 
in the form of the
Gibbs triangle.  The Gibbs phase rule states that
in a system of  $r$ components and $M$ coexistent phases
it is possible arbitrarily to preassign $r-M+2$ variables from
the set $T,P,x^i_j, i=1\cdots M, j=1\cdots r-1$ (Callen (1960)).
In particular, it means that the maximum number
of coexistent phases in the system is $r+2$. One should note here that if 
the system spontaneously orders
in the arbitrarily small external field (magnetic or electric)
then the number of coexistent phases
can be larger 
if we include in the set of thermodynamic variables these fields.

So far we have considered the liquid mixtures in the bulk system. 
The confinement  of the fluid in a capillary also affects mixing
and solubility. The detailed discussion of thermodynamics of
confined mixtures can be found in Evans and Marconi (1987). However, 
the theoretical problem of whether confinement  increases or decreases
solubility and mixing is still not resolved. The problem is especially
important for narrow pores of the size of few molecular diameters.

\noindent{\bf STATISTICAL MECHANICS THEORIES OF MIXTURES}

The modern statistical mechanical
approach to inhomogeneous and/or homogeneous fluids and fluid mixtures is
based on density functional theory (DFT) (Evans (1979)).
The central quantity of interest in DFT 
is the Helmholtz free energy $F[\rho_1\cdots
\rho_m]$ which
is a unique functional of the local 
densities $\rho_i({\bf r})$ ($i=1\cdots m$) in the
$m$ component simple atomic
mixture. In the homogeneous system $\rho_i=N_i/V$ 
where $N_i$ is the number of molecules of type $i$ 
and $V$ is the volume.
The
Helmholtz free energy as a functional of the densities completely specifies
its statistical and thermodynamical properties.
The equilibrium state of the system corresponds to the global minimum
of the grand thermodynamic potential, obtained from $F$ by a
Legendre  transform. 

In the case of anisotropic liquids (e.g.  liquid crystals)
consisting of elongated rigid molecules
one particle distribution function,
$\rho({\bf r}, \omega )$, which depends on the position of the center of
mass ${\bf r}$ of the  molecule and its orientation in space $\omega $,
represented by the
three Euler angles, describes the structure of the
system. Consequently the free energy
of liquid crystals is a functional of $\rho({\bf r}, \omega )$.
More complicated cases are also possible, 
depending on the structure  and flexibility
of molecules. 

For simplicity we 
confine  further discussion to simple atomic fluids.
The free energy can be split into the ideal part 
and the excess part as follows:
$$F[\rho_1\cdots\rho_m]=F_{id}
[\rho_1\cdots\rho_m]+F_{ex}[\rho_1\cdots\rho_m].\eqno(6)$$
For the non-interacting system the excess part is zero and the ideal part 
can be easily calculated from the partition function:
$$F_{id}=k_BT
\sum_{i=1}^m\int 
d{\bf r}\rho_i({\bf r})\{\ln (\rho_i({\bf r})\lambda_i^3)-1\},\eqno(7)$$
where $\lambda_i$ is the thermal de Broglie wavelength of the $i$-th 
component 
in the mixture. The integrals are performed 
over the whole volume of the system.
The excess part, arising from the interactions, 
can be writen in the following general
form:
$$F_{ex}[\rho_1\cdots\rho_m]=\sum_{i,j=1}^m\int d{\bf r}_1d{\bf r}_2
\rho({\bf r}_1)\rho({\bf  r}_2)\int_0^1
d l (l-1)c_{ij}({\bf r}_1{\bf r}_2;[l\rho_1,\cdots ,l\rho_m]).\eqno(8)$$
The functions $c_{ij}$ are the Ornstein Zernicke direct correlation
functions (Hansen and McDonald (1986)). Although this relation is formally
exact, the form of the direct correlation functions is in general 
unknown for 
inhomogeneous and even for  homogeneous liquids. 
Various approximations 
to the exact DFT functional 
has been proposed: weighted density approximation 
(WDA) (Tarazona(1985))
(Curtin and Ashcroft (1985)) and modified weighted density approximaton
MWDA
(Denton and Ashcroft (1989)) for pure systems, later applied to binary
mixtures of hard spheres
(Denton and Ashcroft (1990) (1991))
(for different approach to mixtures see Xu and Baus (1987), (1992)). 
(for review Baus (1990)).
Density Functional Theories have been also applied to the description of
structural phase transitions in liquid crystals
(Poniewierski and Ho\l yst (1988), (1990), Allen et al (1993))).
A recent review on this subject 
is given by L\"owen (1994) and Allen et al (1993).
In general,
the theory has been aimed at explaining the 
properties of inhomogeneous systems from the
known properties of homogeneous systems. 
Usually the form of the direct correlation functions for homogeneous 
system are
required as the input for the theory.
For the homogeneous mixture these functions are defined via the 
Ornstein Zernicke (OZ) equation:
$$h_{ij}(\vert{\bf r}_1-{\bf r}_2\vert)=
c_{ij}(\vert{\bf r}_1-{\bf r}_2\vert)+\sum_{k=1}^m x_k\rho\int d{\bf r}_3
c_{ik}(\vert{\bf r}_1-{\bf r}_3\vert)h_{kj}(\vert{\bf r}_3-{\bf r}_2\vert)\eqno(9)$$
where $\rho=\sum_{i=1}^m N_i/V$ is the total density and $h_{ij}$ is the
two point correlation function. 
This set of equations with suitable closure (e.g.
Percus Yevick (PY) for hard spheres (Henderson and Leonard (1971))  
can be solved  and the functions $c_{ij}$ 
can be calculated. Now combining these results together with MWDA
completely specifies the properties of the inhomogeneous and/or homogeneous
mixture.  

In general, solving the OZ equation for mixtures interacting with
complicated potential is very difficult. One can then resort to some
old methods known in the theory of mixtures as 
a one fluid approximation in the more general scheme known as
conformal theories of mixtures (section 2.3)
(Rowlinson and Swinton (1982), 
Hansen and McDonald (1986)).

\noindent {\bf 2.1 Homogeneous system and ideal mixtures}

In homogeneous mixtures one  defines
mixing functions i.e. for any
thermodynamic functions $A(N_i)$ the mixing function is
defined as
$$A^{mix}=A(N_1\cdots N_m)-\sum_{i=1}^mA_i(N)\eqno(10)$$
where $N=\sum_{i=1}^m N_i$ is the total number of particles and
$A_i$ is the thermodynamic function of interest for the pure
system of $i$-th component. The mixing function vanishes for the
pure system.
If the interactions between particles are the same 
then the form
of the thermodynamic potentials is very simple.
The Helmholtz free energy (same for Gibbs free energy)
is then (Eqs.(6-8)):
$$F^{mix}=Nk_BT\sum_{i=1}^mx_i\ln x_i\eqno(11)$$
Strictly speaking  
real mixtures are non-ideal except in very special cases when 
we have mixtures of isotopes of law  molecular mass 
and when quantum effects are negligible. However, often
the ideal solution is a very good starting point in the
description of many properties of mixtures e.g. gas-liquid 
coexistence.

\noindent{\bf 2.2 Homogeneous systems and excess mixing functions}

The difference between the actual value of the thermodynamic 
mixing function and  its ideal value (Eq.(11)), for the same temperature,
composition and volume (or pressure),
is called the excess
function (the same name is used in DFT (Eq.(8)), but
its meaning there is different). 
The simplest form of the mixing function 
for a binary mixure is given
by the Guggenheim quadratic form 
(Henderson and Leonard (1971), Rowlinson and Swinton  (1982),
$$F^{E}=N\chi x_1x_2\eqno(12)$$
obtained in the simplest lattice approximation. 
In the lattice approximation both the excess Gibbs free energy and 
the excess Helmholtz free energy are the  same and $\chi/k_BT$ 
is the interaction parameter
independent of temperature and pressure (volume).
Lattice models are often used for the description of certain properties
of mixtures (Furman et al (1977), Walker and Vause (1983), Carneiro  and Schick (1988)).
By comparing Eq(10-12) and Eq(6-8)
it follows that in general $\chi$ is a complicated 
integral of  the direct correlation
function. Only in the limit of low density $\chi$ is  independent of
$x_i$. In this limit $c_{ij}=f_{ij}=\exp{(-v_{ij}/k_BT)}-1$, where
$v_{ij}(\vert{\bf r}\vert)$ is the two body interaction potential between
molecules of components $i$ and $j$. In this  approximation
$\chi$ can be explicitely calculated from
Eq(8). 

Usually in thermodynamics one defines by Eq(12) the Gibbs free energy,
$G^{E}$,
assuming that $\chi$ is the function of temperature and pressure.
Then the heat of mixing (enthalpy)
$$H^{E}=N\left(\chi-T{{\partial \chi}\over{\partial T}}\right)
x_1x_2\eqno(13)$$
and the change of volume in the system upon mixing is
$$V^{E}=Nx_1x_2{{\partial \chi}\over{\partial P}}\eqno(14)$$
Most mixtures of simple liquids (e.g. argon, krypton)
have positive heat of mixing i.e. heat is absorbed, but the excess volume
$V^{E}$ can be of either sign. The excess quantities are usually small,
e.g. $V^{E}$ per mole in the 50\%  solution 
(in mole fraction) of  tetrachloroethene  in
cyclopentane at $25^\circ$C
is -8 mm$^3$ (the mixture contracts) 
(total volume is $\sim 10$cm$^3$) 
and the excess heat of mixing 
$H^{E}$ is
$800 J$ per mole.

\noindent{\bf 2.3 Conformal fluid theories and one fluid approximation}

One of the oldest approaches to the theory of homogeneous 
mixtures dates back to van der
Waals and it is known as the one fluid approximation.
In this approach the excess properties of the mixture are expressed in
terms of the quantities of a hypothetical pure fluid.
It applies to mixture in which
the interaction potential $v_{ij}(r)$ satisfies  the scaling relation
$$v_{ij}(r)=\epsilon_{ij} v(r/\sigma_{ij})\eqno(15)$$
for all $i,j$. Often $\sigma_{ij}$ and $\epsilon_{ij}$ (for $i\ne j$)
are related to pure components. The most frequently used combination rules
are as follows: $\sigma_{ij}=(\sigma_{ii}+\sigma_{jj})/2$ and
$\epsilon_{ij}=\sqrt{\epsilon_{ii}\epsilon_{jj}}$
 known as the Lorentz and
Berthelot rule , respectively.
The hypothetical fluid is characterized by the potential given by Eq(15)
with some energy parameter $\epsilon$ and length parameter $\sigma$.
In order to specify these parameters
one makes the following approximation:
$$h_{ij}(r)=h(r/\sigma)\eqno(16)$$
and sets
$$\sigma^3=\sum_{i,j=1}^m  x_ix_j\sigma_{ij}^3\eqno(17)$$
The interaction parameter for this hypothetical one component fluid
reads
$$\epsilon=\sum_{ij=1}^m x_ix_j\epsilon_{ij}{{\sigma_{ij}^3}\over{\sigma^3}}
\eqno(18)$$
and the interaction potential is $\epsilon v(r/\sigma)$. Now the excess mixing
properties of the mixture are related to the 
excess properties of this fluid.
This approximation is known as the van der Waals one-fluid approximation
(Henderson and Leonard (1971)). 
The comparison with computer simulations
for the model binary mixture of Lenard-Jones fluids shows a very good
agreement with the theory (Hansen and McDonald (1986)). The 
approximation breaks down when the molecules of different components are
very disparate in sizes and have 
very different energy parameters. 
Another of its drawbacks is that it always leads to
a negative volume change , $V^{E}$, upon mixing.
More  complicated approximations
are also possible but usually lead to worse agreement
with computer simulations. For real systems the simple Lenard Jones mixture
constitutes a 
rather poor approximation (Hansen and McDonald (1986)).

One has to  remember that excess mixing quantities are rather
small and that small deviations of the model interaction potential
from the real potential between the molecules can change even the sign of
the excess quantity (e.g. $V^{E}$).

\noindent{\bf 3. INTERACTION POTENTIAL AND MIXING}

The intermolecular forces have their origin in quantum mechanics and
classical electrostatics. According to  the Hellman-Feynmann theorem
the interactions between the molecules can be  calculated according
to the laws of electrostatics, once the distribution  of  electrons 
has been determined from the Schr\"odinger equation. 
>From the simple Coulomb potential originates various important interactions
at the molecular level.
The dispersion
van der Waals attractive interactions are
responsible for boiling  (gas liquid transition). 
It is now known that the reduction of the range of
attraction in the  system results in the complete disappearence of the
liquid phase even if the attraction is strong at short distances
(Tejero et al (1994)). It implies for example that $C_{60}$,
where interactions are of the short range, is a substance
which should not have a liquid phase (Hagen et al (1993)).
Molecules, when brought together, strongly repell each other, preventing
the overlap of electronic orbitals. This interaction gives the molecule
its size and shape and is 
known as hard core repulsion or steric interaction.
Steric interactions  
are responsible for freezing transition (Alder and
Wainwright (1957))
structure of liquids (Hansen and McDonald (1986)), 
and structural transitions
in liquid crystals (Onsager (1949), Frenkel (1990)).
Van der Waals attraction and steric repulsion
are very important since they are
present in every molecular system.
Apart from them there are many specific interactions.
One interesting example considered here is
hydrogen bonding occuring between e.g. water molecules.
This interaction is responsible for the fact that ice has lower density
than liquid water; and that the highest density of water occurs at
4$^\circ$C. The extremely high boiling and freezing point of water is also due to
this interaction and the tetrahedral structure formed 
by water molecules.
Here we shall try to partially answer the question how these various 
interactions
affect mixing. 

\noindent{\bf 3.1 Van der Waals interactions}

If we apply  the Berthelot rule, which holds resonably well for van der Waals 
forces (Israelachvili (1985)), in the Guggenheim model (Eq(12))
for the binary mixture we find:
$$\chi\sim(\sqrt{\epsilon_{11}}-\sqrt{\epsilon_{22}})^2>0.\eqno(19)$$
Since $\chi$ is always greater than zero, 
we may conclude that van der Waals
interactions promote demixing. At sufficiently low temperatures
the excess part of the Gibbs free energy outweights the ideal part (Eq.(11))
and demixing occurs. With this simple model we obtain the type of phase 
diagram shown in Fig.5. In the first approximation the square root of the
energies $\epsilon_{ii}$ are proportional to the 
polarizability of the 
molecules of $i$-th component.

The van der Waals interactions for any two bodies in vacuum
are always attractive. Also they are  attractive for 
identical bodies in  a solvent. However, one may have repulsive
van der Waals interactions between two different solute
molecules in a solvent. This happens whenever the
index of refraction (related to polarizabilities)  of the solvent is  
intermediate between two different solute molecules (Israelachvili (1985)). 
It happens in many mixtures of organic solvents and different organic 
polymers (Van Oss et al (1980)).

\noindent{\bf 3.2 Steric interactions}

It has been believed until recently that steric interactions alone
cannot induce demixing. This belief has been based on the fact that
the ideal entropy of mixing  
decreases if the demixing transition takes place.  
This decrease of the entropy, $S$, 
can be compensated by the decrease of the energy, $U$, 
so that the free energy, $F=U-TS$ decrease upon mixing.
Since in  the hard core system the interaction energy is zero, one 
has not
expected the demixing transition in such a system.
This conclusion is supported also by the explicit calculation
of $\chi$ parameter (Eq.(12)) for the binary fluid mixture of hard
spheres in the low density approximation. One finds $\chi <0$
in this limit and  concludes that mixing  is favored in the system.
 
Recent computer simulations have shown that this is not the case and  that
in a binary mixture of model system of hard core molecules differing
in size only the demixing transition driven by the increase of entropy
occurs.
Although the ideal entropy is the largest in a
homogeneous mixture, the entropy associated with the 
free volume is larger in the demixed fluid ( Biben and Hansen (1991)
Frenkel and Louis (1992),
Van Duijneveldt and Lekkerkerker (1993), 
Dijkstra and Frenkel (1994), Dijkstra et al (1994)).
At certain concentration  of solute molecules the latter outweights the
former and demixing occurs. The role of steric interactions in real 
mixtures has not been resolved yet. Clearly it is a many body effect.

\noindent{\bf 3.3 Hydrogen bonding}

The unique properties of water follow from 
its ability to form 
the tetrahedral  structure in the liquid, induced by hydrogen bonding
(Israelachvili (1985)). The interaction is electrostatic, but
its strength ( 10-40  kJ/mol) is one order of magnitude larger than
the strength of  the van der Waals energy ($\sim 1$kJ/mol).
This interaction depends on the orientation of molecules and in general
is not pair-wise additive. 

The solubilization of molecules which do not form hydrogen bonds (e.g.
alkanes and other 
hydrocarbons) in a hydrogen bonding solvent (methane, water, acetic acid)
involves forming a clathrate cage around the solute molecule (Fig.11).
The solvent molecules adapt such orientation close to the  non bonding
solute as to saturate its hydrogen bonds with other solvent molecules.
It means that solvent molecules around the solute molecules are
more ordered than the molecules in the bulk solvent. 
It leads to the decrease of entropy in the process of solubilization, which
for this reason becomes unfavorable.
For n-butane in water the heat of mixing is $\Delta H=-4.2$kJ/mol,
while the change  of the entropy $-T\Delta S$ (at $T=25^\circ$C) is
$28.7$kJ/mol.  Thus, the change of entropy contributes $85$\%
to the change of the Gibbs free energy$\Delta G=\Delta H-T\Delta S$.
For longer hydrocarbons the contribution of entropy to the
total change of the Gibbs free energy is even larger (Israelachvili (1985)).
Low solubility of non hydrogen bonding molecules in water
is entropic in nature and is known as a hydrophobic effect.
The solubility of hydrocarbons in hydrogen bonding
solvent (e.g. methane) decreases with the length of the hydrocarbon, e.g.
the critical temperature for n-pentane is $T_c=287^\circ$C, while for
n-hexane becomes $T_c=307^\circ$C (at normal pressure) (Rowlinson and Swinton
(1982)). 

Methane and water mix very well  in all proportions at room temperature
and this is due to the fact that both form  hydrogen bonds. 
We expect that molecules of right geometry,
containing electronegative atoms (oxygen atoms in
alcohols, nitrogen atoms in amines) are capable of forming 
hydrogen bonds. 
At low temperatures the hydrogen bonds forming between the 
solute and solvent
molecules enhance mixing ($\chi <0$). 
If we raise the temperature the bonds are broken and
liquids demix above the lower critical temperature
due to van der Waals forces  ($\chi >0$). Thus, 
hydrogen bonds are responsible
for enhanced mixing at low temperatures  and
consequently for the existence of 
the lower critical point (Fig.6) (Walker and Vause (1983)).
We expect that this phenomenon should 
strongly depend on the molecule geometry.

\noindent{\bf 4. POLYMER BLENDS AND POLYMERS IN SOLUTIONS}

The simple lattice model presented in section 2 (Eq.(12))
has been applied  to polymer
blends and polymers in solutions by Flory and 
Huggins.
(Flory (1953)). The free energy for the binary mixture of two homopolymers
A and B consiting of $N_A$ and $N_B$ monomers respectively
is given by the Flory-Huggins expression: 
$$F^{mix}/(Mk_BT)={{x_A}\over{N_A}}\ln{x_A}+{{x_B}\over{N_B}}\ln{x_B}
+{{\chi}\over{k_B T}} x_A x_B.\eqno(20)$$
Here $x_A=n_AN_A/(n_AN_A+n_BN_B)$ is the mole  fraction of A monomers
in the mixture,
$n_i$ is the number of polymer molecules of $i$ (A or B) component,
$M=n_AN_A+n_BN_B$ is the total number of monomers in the system and
$\chi$,  
in the polymer physics and chemistry, is 
called the Flory-Huggins interaction parameter.
In the Flory-Huggins theory incompressibility is assumed, same as in 
the Guggenheim model of simple mixtures.
The critical temperature  is
$$T_c={{2N_AN_B}\over{k_B\chi (\sqrt{N_A}+\sqrt{N_B})^2}}\eqno(21)$$
and at the critical composition
$$x_A^c={{\sqrt{N_B}}\over{\sqrt{N_A}+\sqrt{N_B}}}.\eqno(22)$$
The ideal  mixing entropy is reduced by a factor $N_A$ and/or $N_B$ 
and thus, the critical
temperature is very large for the polymer mixtures ($N_A,N_B\gg 1$).
Indeed polymers separate very easily.
Even small differences 
in the interaction potential get magnified by a factor of $N_A,N_B$
and thus the critical temperatures are often much higher then the 
temperature at which the polymer molecule breaks due to the breakage of
chemical bonds linking the monomers. 
Even a  binary mixture of 
isotope polymers (one  deuterated) can undergo a  demixing transition
at  
room temperature if $N_A,N_B$ are sufficiently large (Gehlsen et al (1992)).

A mesoscopic theory of polymer chains has been formulated by Edwards (1966)
and applied to polymer blends,  polymers in solutions, membranes in solutions
etc.

\noindent{\bf 4.1 Polymer blends}

Despite the fact that the Flory-Huggins theory has been used for 
more than four decades, only recently it has been carefully checked
experimentally (Bates et al (1988), Gehlsen et al (1992)) 
and in computer simulations  (Deutsch and Binder (1992)).
It follows, from computer simulations
that $\chi$ parameter has the following form
$${{\chi}\over{k_B T}}={{\alpha}\over T}+\beta,\eqno(23)$$
where $\alpha$ and $\beta$ are constants independent of
temperature or composition. In the first approximation 
$\alpha$ is related to the attractive van der Waals forces while $\beta $
to steric interactions  (called sometimes excluded volume interactions).
Sometimes, in experimental works,
it is assumed  
that $\alpha$ depends linearly on the concentration $x$
(Roe and Zin (1984)).
Typically we find $\alpha\sim 1$ and $\beta\sim 10^{-3}-10^{-4}$.
>From the previous section we know that the effective 
interaction  $\chi$ 
is related to the integrals of the direct correlation
functions (Schweizer and Curro (1988)) and thus 
$\chi$ can be, in principle, a complicated
function of  temperature, density and concentrations. In general,
$\chi$ should not depend on the global architecture  of 
the polymer chain.

In polymer systems the quantity of interest is the radius of gyration, 
i.e. the linear size of  the
region occupied by a single polymer molecule, consisting of $N$ monomers 
of length
$l$. It is defined as follows (Doi and Edwards (1986):
$$R^2={1\over {N^2}}\sum_{i=1}^{N}\sum_{j=1}^{N}\left<\left
({\bf r}_i-{\bf r}_j\right)^2\right>.\eqno(24)$$
Here ${\bf r}_i$ gives  the location of the $i$-th monomer in the polymer
chain and $<\cdots >$ denotes the statistical average.
For the polymer blend the dependence of $R$ on $N$ is
(the leading term in $N$) as follows:
$$R=\sqrt{N{{l^2}\over 6}}\eqno(25)$$
characteristic for the non-interacting chain. 
It means that interactions between monomers in
the same  polymer chain are screened by the presence of other chains
(Doi and Edwards (1986), de Gennes (1979))
or a solvent
(Edwards (1975)). 
As the temperature is lowered the scaling of $R$
with $N$ does not change. However, the chains shrink progressively
($R$ decreases)
as the critical temperature is approached (Ho{\l}yst and Vilgis (1994)).

One  can deduce the structure of a polymer blend from the  scattering
experiments. The scattering intensity $S(q)$ has the following  form
(de Gennes (1979)):
$$S^{-1}(q)={1\over{x_A g(N_A,q)}}+{1\over{x_B g(N_B,q)}}-2{{\chi}
\over{k_BT}},\eqno(26)$$
where  $g(N,q)=2N(y+\exp (-y)-1)/y^2$ and $y=N q^2l^2/6$. The contrast for the
neutron scattering is achieved by the deuteration of one of the components.
This approximation is known as the random phase approximation (RPA).
The characteristic length in the problem is given by  the radius of gyration
$R$ (Eq(25)).

>From the theoretical point of view it is interesting to note that
although
the critical point in a 
binary mixture of polymers belongs to the universality
class of the Ising model, in the limit $N\to\infty$ the mean field
theory becomes exact. In particular the critical exponents
measured for the system have the mean field values for 
$\vert T-T_c\vert /T_c\ge 1/N$ (de Gennes (1977), 
Ho\l yst and Vilgis (1993)). 
This behavior has been confirmed 
in experiments (Schwahm et al (1987)). 

Polymer blends are compressible mixtures (Floudas et al (1993)), 
but the role of this factor in the Flory-Huggins theory 
is still under theoretical study (Lifschitz et al (1994)).

\noindent{\bf 4.2 Polymers in solutions}

In the theory and experiment one often discusses 
three types of solvents: good, bad and theta solvents.
Since for $N_A\gg N_B=1$ the critical concentration is very low 
($x_A^c\sim\sqrt{1/N_A}$) we can
expand $F^{mix}$ in $x_A$  and find for this dilute solution:
$$F^{mix}/(Mk_BT)=
{{x_A}\over{N_A}}\ln{x_A}+
B x_A^2+{1\over 6}x_A^3\cdots
,\eqno(27)$$
where $B=({1/2}-{{\chi}/{(k_B T)}})$ is the osmotic virial coefficient
(Atkins (1993), de Gennes (1979)). For $B>0$ the solution is classified as
bad, for $B<0$ it is classified as good and for $B=0$ it is classified as
theta or Flory solvent. The nature of the solvent depends
on temperature. For each solvent there is a unique temperature 
(called the theta temperature) when
$B=0$ and the solution becomes nearly ideal. In  a theta solvent 
the chains behave, in the first approximation,  
according to Eq.(25). 
In a good solvent the
polymer molecule swells and
$$R\sim N_A^\nu\eqno(28)$$
with the Flory exponent $\nu\approx 3/5$, which should be compared with
$\nu=1/2$ (Eq(22)) for the polymer chain in a blend and polymer chain in a
theta solution. In a bad solvent the chains shrink and $\nu\le 1/2$.
In fact, close enough 
to the critical temperature each solvent becomes a bad solvent.
The Flory-Huggins theory predicts correctly the
location of the critical point (de Gennes (1979), Ho\l yst and Vilgis (1993))
but not the shape of the coexistence curve close to the critical point.
More informations on polymers in solutions can be obtained in a recent
monograph (des Cloizeaux and Jannik (1990)).

It is often assumed that the theta temperature must reflect the
competition between the attractive van der Waals forces and 
the short range steric interactions. However, it does not have to be
the case.
As shown explicitely by Frenkel and Louis (1992) in a simple
lattice model of a polymer chain consiting of monomers 
interacting with steric interactions only and
solvent consisting of small molecules interacting with the 
the same steric interaction the Flory-Huggins parameter is
$${{\chi}\over{k_BT}}={1\over 2}c\ln{(1+z_s)}>0.\eqno(29)$$
Here $c$ is a coordination number of the lattice and $z_s$ is the
fugacity of the solvent molecules.  Thus, in this system with
purely steric interactions we may have a bad, good and theta solvent
by changing the chemical potential of the solvent (Dijkstra et al (1994)).
The mechanism of demixing in such a mixture is clear. The fraction of
the 
volume accesible to small  solvent particles increases when the large 
particles cluster. Of course, in this model  system, demixing is a purely
entropic effect.
Also the coagulation (clustering) of colloidal particles
in a polymer solution is 
an entropic effect (Meijer and Frenkel (1994)).
The clustering of colloidal particles increases  the total number
of accesible polymer configurations. The effective attraction between
the colloidal particles, responsible for coagulation, is not
pair-wise additive (Shaw and  Thirumalai (1991)) and that is why
its quantitative theoretical description is so difficult.
The role of steric interactions in polymer solutions is 
not fully understood.

We note that in the process of preparation of the polymer blend,
one first prepares a solution of A and B polymers in a solvent and then
evaporates the solvent. The comparison of the theory and experiment
is often plagued by the inevitable polydispersity of the
polymer masses.
And finally, 
we point out that often instead of using mole  fraction one uses
either volume fraction or simply the density of monomers 
in the description of the free  energy of mixing in polymer solutions
(Atkins (1993), De Gennes(1979)).

As we see, polymers are not easily mixed even at high temperatures. 
Usually chemical means are used to induce mixing. One 
of them is  a chemical modification of one of the components
in such a way as to induce hydrogen bonds between the different
components. For example  PSD (deuterated polystyrene)  and PBMA
poly (butyl  methacrylate) are not miscible. However,  modifying
PSD by attaching at random OH groups along the chain makes 
PSD(OH)/ PBMA blend miscible. It is due to the fact that 
OH groups from PSD(OH) form hydrogen bonds with CO groups
on any segment of PBMA (Hobbie et al (1994)).
Another way to induce miscibility is  even more direct. One can simply 
join one polymer chain with the other by chemical methods. In this way
diblock and multi block copolymers are formed.

\noindent{\bf 5. ORDERING AND DEMIXING}

There are many examples of demixing induced by 
ordering and vice versa.
These effects occur in  complex fluids such as
diblock copolymers (Bates (1991), 
Bates and  Fredrickson (1990)), amphiphilic systems
(Gompper and Schick (1994),  Laughlin (1994)), and
mixtures of liquid crystals (De Gennes and Prost (1993)).

\noindent{\bf 5.1 Mixtures of liquid crystals}

For commercial applications in the display industry
only mixtures of liquid crystals are used. 
By mixing two nematic liquid crystals
one can lower its freezing point (eutectic composition Sec 1.
on Thermodynamics)
without strongly affecting its isotropic-nematic phase transition point.
Such  mixture exhibits anisotropic properties in the liquid state in a much
wider temperature range than the pure substances. Usually the point at which
the nematic-isotropic transition occurs is called the clearing point.  

Miscibility is used as a first test for new
phases in liquid crystals. If the two materials  give the same texture 
and are 
miscible in all proportions maintaining this texture they must have the same
symmetry. However, we have learned that demixing does  not have to reflect
differences in the  
symmetry of phases.  Differences in the shape and size of molecules
are sufficient to induce demixing. Nonetheless, the miscibility test
is still very useful, since it is quick and simple.

NMR allows to measure internuclear distances of solute molecules
if the solvent is nematic. 
Elongated isomers are more readily to be solubilized by
nematic solvents and this effect is used in chromatography.

Doping
the nematic solvent with chiral solute induces a cholesteric (twisted)
ordering in the system with a very large pitch (the wavelength
of the twist). Changing the composition
of this mixture allows to vary the pitch continuously. 
The pitch, in a dilute
solution, is inversely proportional to the concentration of the
solute molecules. In chiral, ferroelectric 
smectic C$^*$ one can increase the
pitch by dissolving in the system non chiral solute.

In the homogeneous solution of the small monomeric units and nematic 
solvent  the fast polymerization reaction between the monomers results
in the demixing 
and leads to the formation of the polymer dispersed liquid crystals
(PDLC), which are used as switchable windows, light shutters, displays.
Demixing is induced here 
since the length of the polymeric component of the mixture
grows in the polimerization reaction and according to Eq.(21) the critical
temperature gets higher for longer chains. If the reaction is fast, the 
droplets of the nematic solvent, which forms in the process of demixing, 
are homogeneously distributed in
the solid polymeric matrix. Usually the reactions require more components
such as curing agents, catalysts etc (Doane (1990)). 

The extremely rich phase behavior 
of liquid crystals and liquid crystalline polymers
has been described theoretically 
in several models: lattice spin model ( Sivardi\'ere (1980)),
Onsager model for elongated molecules (Deblieck and Lekkerkerker (1980),
Van Roij and Mulder (1994)), combination of the Flory-Huggins model for mixtures
and Maier-Saupe model for liquid crystals (Brochard et al (1984)),
Ma\"issa and Sixou (1989), Ho\l yst and Schick (1992)).
The following results follow from these theoretical studies.
Nematic-isotropic phase transition in mixtures is always accompanied
by partial demixing. Nematic phase is always richer in the
component which orders easier. Azeotrope 
at the nematic-isotropic phase transiton is found in binary 
mixtures where both components have
very similar isotropic-nematic transition temperatures
in pure systems. 
The azeotrope concentration is very close to the 
critical concentration for demixing inside the nematic phase.
All the diagrams shown
in (Figs.(1-10)) are found in binary mixtures of liquid crystals if 
instead of vapor we have the isotropic phase and 
instead of liquid the nematic phase.
The topology of many diagrams 
is even more complicated (Brochard et al (1984)).
Nematic order parameter (De Gennes and Prost (1993)) always couples 
to
the concentration.
The demixing at the isotropic nematic phase transition is driven by ordering,
however, inside the nematic phase we can also observe 
the isotropic demixing (Fig.5,9) which changes the nematic order parameter.
The demixing inside the ordered nematic mixture has the same origin 
(isotropic van der Waals or steric interactions) as the
demixing shown in Fig.5 for simple isotropic liquid mixture. 
The roles of the anisotropic steric (shape) interactions and 
attractive interactions in the process are not fully understood.
The demixing in the ordered nematic  liquid mixture (Casagrande et al (1982),
Dorgan and Soane (1990)) and ordered 
smectic liquid 
mixture (Sigaud et al (1990)) has been  observed in experiments,
confirming many of the above stated predictions.
 
\noindent{\bf 5.2 Diblock copolymers}

Polymers are usually not miscible (section 3.), but at the same time
for practical applications we need homogeneous mixtures. One way to prevent
demixing transition is the chemical bonding of A polymer to the B polymer.
In this way diblock and 
multiblock copolymers are formed (Bates and Fredrickson (1990)).
Although the phase 
separation is now prohibited at the macroscopic scale, the system
can undergo the so called microphase separation. The `demixing'
in the AB diblock copolymer system takes place at
the scale given by the size of the  radius of gyration and at 
much lower temperatures
than the demixing in the polymer AB blend. 
The system forms many ordered phases
in a liquid state: lamellar, gyroid (Matsen and Schick (1994)), 
diamond, cubic and
hexagonal. The microscopic interactions responsible for the formation
of these phases are the same as those responsible for the demixing transition
in the AB polymer blend (Leibler (1980))
and the same Flory-Huggins parameter $\chi$ is used
in the description of both systems. The AB diblock copolymer
molecules form A rich and B rich domains and
chemical bonds joining A and B 
molecules reside at the interface between the domains (Fig.12).
The molecules are stretched in the ordered phase at low temperature
and the characteristic period of the ordered structures scales 
with the number of monomers in a molecule, $N$, as $N^{2/3}$.
The mixing of the A and B homopolymers can be enhanced by the 
addition of the AB diblock  copolymer. The latter has 
similar effect on the
AB polymer blend
as an amphiphilic molecule on the mixture of oil and water,
although amphiphiles can be more effective in enhancing mixing
between immiscible liquids
(Ho\l yst and Schick (1992)).

\noindent{\bf 5.3 Amphiphilic systems}

The amphiphilic molecules consist of two parts:
a polar head which can form hydrogen
bonds and a hydrocarbon tail. Such molecules are not easily solubilized
in water because of their long tails.  
At a certain, usually low,
concentration called
critical micelle concentrations the amphiphilic molecules
aggregate and form closed structures called micelles.  The interior of
the micelle consists of the hydrocarbon tails
and has the structure and density of the hydrocarbon liquid. The polar
heads reside on the surface of the micelle, shielding solvent molecules
(water) from the interior hydrocarbon liquid. 
The geometrical shape of the micelles
depends on the maximum length
of the hydrocarbon chain, $l_c$, cross section area occupied by the 
polar head, $a_0$,
and hydrocarbon volume,$v$ (Israelashvili (1985), Proceedings (1985)). 
For example, spherical micelles can form for $v/(a_0l_c)<1/3$, nonspherical
micelles (in particular cylindrical) for $1/3<v/(a_0 l_c)<1/2$ and
flat bilayers or vesicles for $1/2<v/(a_0l_c)<1$. They are
usually polydisperse in size. 
As we see, the hydrogen bonding and hydrophobic effect are responsible 
for the formation of
these structures.
When the concentration of amphiphilic molecules
becomes comparable to the concentration of the water solvent,
certain ordered, bincontinuous phases are formed
such as:  
simple cubic (Fig.13), diamond (Fig.14),  
and  gyroid structures (Fig.15).
The internal surfaces of these structures 
are formed by the polar heads. The structures are periodic
in three dimensions and the internal surfaces
may assume the configuration
of minimal surfaces. 
The minimal surface is characterized by zero mean curvature
at every point;
each point on this surface is a saddle point.

Amphiphilic molecules strongly reduce the surface tension between 
hydrocarbon liquid and water and enhance mixing between these two immiscible
liquids. One characteristic structure formed in the ternary mixture of
oil, amphiphiles and water is the microemulsion. This homogeneous
phase is characterized experimentally by the enhanced scattering
at a certain nonzero $q$ vector. The form of the 
water-water scattering intensity
$$S(q)=(a+gq^2+cq^4)^{-1}\eqno(30)$$
with negative $g$ fits the data from scattering experiments very well
(Teubner and Strey (1987)). It reveals the internal structure of the 
microemulsion
with a characteristic length of order of 400\AA (size of the amphiphile 
molecule is about 30\AA). 
Inside the microemulsion water rich regions
of the size of the characteristic length are separated by the
amphiphiles from the oil rich regions of a similar size. 
On the macroscopic scale the 
system can be characterized by the large amount
of these internal water-oil interfaces.
Various surface and bulk properties of microemulsions are
described by Ciach (1992)  and Gompper and Schick (1994).
The simplest Landau model which can be used for the description of
bulk and surface properties of
microemulsion is given by the following functional:
$$F[x({\bf r})]=\int d{\bf r} ((\triangle x)^2+
g(x)(\nabla x)^2+f(x)),\eqno(31)$$
where, $\triangle$ denotes the laplacian, $\nabla$ the gradient,
$$g(x)=g_2x^2-g_0\eqno(32)$$
and
$$f(x)=x^2(x-1)^2(x+1)^2.\eqno(33)$$
Here x is the local concentration difference
between oil and water, $f(x)$ is the simplest Landau bulk
free energy for the three phases (pure water, 
pure oil and microemulsion) 
at coexistence and $g_2$ and $g_0$ are constants. 
Interestingly the bicountinuous phases shown in
Figs. 13,14,15. correspond to the local minima of this functional.

\noindent{\bf 6. KINETICS OF DEMIXING} 

When the homogeneous binary 
AB mixture above its critical point is suddendly 
cooled (quenched) below its critical temperature it ceases to be 
in the thermodynamical equilibrium
(for review on kinetics see Langer (1992)). 
The homogeneous state
can be now either a metastable or unstable state. In the case 
of metastable state the process of demixing requires, in the first
place, nucleation of droplets of the minority phase, say A rich phase.
Then the droplets starts to grow. At first they grow independetly
and their radius, $L(t)$ changes with time,$t$, 
according to the formula:
$$L(t)\sim \sqrt{t}\eqno(34)$$
This behavior has been observed in binary polymer blends (Cumming et al
(1990)).
At the later stage significant fraction of
A molecules disappear from the homogeneous mixture (most of them form
droplets) and competitive growth starts. Small droplets 
decrease in size and the A atoms diffuse towards large droplets which
grow. This mechanism is known as `evaporation-condensation'
mechanism and is described by the 
Lifshitz-Slyozov-Wagner (LSW)
law for the growth rate of large droplets:
$$L(t)\sim t^{1/3}\eqno(35)$$
When  the system is quenched into the thermodynamically unstable region
the demixing proceeds via the spinodal decomposition mechanism. 
Early stages of this process are described by the Cahn-Hilliard theory.
According to the theory the system 
becomes unstable with respect to small fluctuations
of wavevector $q$  smaller than some value $q_0$. 
The key prediction of the theory is
the exponential growth of the scattering intensity $S(q,t)$ in time with
a well defined maximum at $q_{max}=q_0/\sqrt{2}$.
Interpenetrating 
A rich and B rich domains of the size of  $L\sim 1/q_{max}$ are
formed.  
The mixtures, of low viscosity, undergoing spinodal decomposition 
do not remain 
for a long time in their unstable
early configuration, contrary 
to the high molecular weights polymer blends. The
latter are very viscous liquids and  the whole kinetics of phase
separation is very slow allowing a detailed experimental
observation of the process (Bates and Wiltzius (1989)).
In the late stages of spinodal decomposition the scattering intensity
can be represented by the following scaling form:
$$S(q,t)\approx L^d(t)Y(qL(t))\times {\rm Const}\eqno(36)$$
where $Y$ is a scaling function, $d$ is the dimension of  space and 
$L(t)$ is the time dependent length characterizing
the size of the interpenetrating 
A-rich and B-rich regions when the volume fractions are equal. 
$L(t)$ grows in time,  the pattern coarsens and the interfacial
area, whose energy drives the coarsening, decreases. In the
case of the unequal volume fractions $L(t)$ is
the size of A (minority phase) droplets (Eq.(35)).
The late stage configuration depends on the initial volume fractions, but
not on the early stage mechanism of demixing. 
In the late stage we have either a dilute gas of
droplets of the minority phase in the sea of the majority phase,
dense system of the droplets of one of the phase in the sea of another
or the
interpenetrating network of A-rich and B-rich domains which coarsen in time.
The growth rate and the growth 
mechanism for the first case is described by the
LSW law(Eq.(35)). 
In the second case a Binder-Stauffer mechanism of collisions and 
coalescence of droplets is valid. It gives $L(t)\sim t^{1/3}$, similarly
as in the first case. Finally, in the last case
if the coarsening proceeds via the flow induced by the surface tension
of interfaces, the scaling law $L(t)\sim t$ holds.
The scaling function $Y$ and the late stage of spinodal decomposition 
have been
studied in computer simulations (Koga and Kawasaki (1993)).
Although the field of kinetics of demixing or, in general, of 
first order 
phase transitions is rather old, new experimental results suggest that
the problem is far from being understood.
First of all, a new mechanism of coarsening 
called `collision induced collision' has been observed 
(Tanaka (1994)). It has been also shown that surface effects 
and confinement play an important 
role in spinodal decomposition (Tanaka (1993), Jones et al (1991),
Wiltzius and Cumming (1991)).
Finally, the problem of heat release  during demixing has been addressed
in experimental studies (Bailey and Cannell (1993)).
Concluding: the kinetics of demixing is still an active field of research.

\noindent{\bf GLOSSARY}

\item{}  {\bf Solubility}:
is the ability of the substance to form a solution with another substance.
It also denotes the maximum amount of the solute  that can be solubilized
in the solvent at the given thermodynamic  conditions.
\item{} {\bf Solution}: 
is a single homogeneous liquid, solid or gas phase that is a 
mixture in which the components
(liquid, gas, solid or  the combination thereof) 
are uniformly distributed throughout the mixture. 
\item{} {\bf Miscibility}:
denotes the tendency or capacity of two or more 
liquids to form a uniform blend,
that is to dissolve in each other. 
\item{} {\bf Solute}:
the substance
less abundant in the mixture.
\item{} {\bf Solvent}: the substance 
most abundant in a mixture. 
\item{}  {\bf Upper (Lower)  Consolute (Or Critical) Temperature
For The Liquid Mixture}:
temperature, for the the binary mixture of A and B components,
above (below) which A and B components mix in all proportions i.e.
they exhibit full miscibility.
\item{} {\bf Azeotrope}: the point on the phase diagram where
two different phases of same composition coexist.
Usually in simple liquids the two phases are: vapor and liquid.
\item{} {\bf Eutectic Composition}: the liquid mixture composition
with the lowest freezing point.
\item{} {\bf Excess Thermodynamic Quantities}: the difference between
actual mixing functions and ideal mixing functions.
\item{} {\bf Mixing Functions}: the difference between the function 
defined for the mixture and the sum of these functions 
for pure systems. The mixing function vanishes for the pure system.
(Eq(10)).
\item{} {\bf Ideal Mixture}: mixture in which all the components mix
in all proportions without the change of volume or enthalpy.
Interactions between various components in the mixture are the same.
See also Eq(11)).
\item{} {\bf Spinodal Decomposition}: process of phase separation  
of the mixture in
the region (of the phase diagram)
of the thermodynamical instability of the mixture.

\noindent{\bf Works  Cited}

\item{} Alder, B.J. and Wainwright, T.E., 
(1957) {\it J.Chem.Phys.} {\bf 27}, 1208-1209.
\item{} Allen, M.P., Evans, G.T., Frenkel, D. and Mulder, B.M. (1933),
{\it Advances in Chemical Physics}, 
Vol. XXXVI, ed. Prigogine, I. and Rice, S.A.
John Wiley \& Sons, Inc, 1-166.
\item{} Atkins, P.W., (1993) {\it Physical Chemistry}, 
Oxford University Press.
\item{} Bailey A.E. and Cannell, D.S. (1993), {\it Phys.Rev.Lett},
{\bf 70}, 2110-2113.
\item{} Bates, F.S. (1991), {\it Science}, {\bf 251}, 898-905.
\item{} Bates, F.S. and Fredrickson, G.H. (1990), {\it Annu.Rev.Phys.Chem.}
{\bf 41}, 525-557.
\item{} Bates, F.S. and  Wiltzius, P. (1989), {\it J.Chem.Phys.},
{\bf 91} 3258-3274.
\item{} Bates, F.S., Muthukumar,M., Wignall, G.D. and Fetters, L.J. (1988),
\hfill\break {\it J.Chem.Phys.} {\bf 89}, 535-544.
\item{} Baus, M. (1990) {\it J.Phys.Condens.Matter} {\bf 2}, 2111-2126. 
\item{} Biben, T. and Hansen, J.P. (1991), {\it Phys.Rev.Lett}, {\bf 66}, 
2215-2218.
\item{} Brochard, F., Jouffroy, J. and Levinson, P. (1984),
{\it J. de Physique}, {\bf 45}, 1125-1136.
\item{} Callen, H.B. (1960), {\it Thermodynamics}, John Wiley \& Sons, Inc.
\item{} Carneiro, G.M. and Schick, M. (1988), {\it J.Chem.Phys.} 
{\bf 89}, 4368-4373.
\item{} Casagrande, C., Veyssi\'e, M and Finkelmann, H. (1982),
{\it J.Physique Letters}, {\bf 43}, L671-L675.
\item{} Ciach, A. (1992) {\it Polish Journal of Chemistry}, {\bf 66}, 1347-1381.
\item{} Cumming, A., Wiltzius, P. and Bates, F.S. (1990),
{\it Phys.Rev.Lett.}, {\bf 65}, 863-866.
\item{} Curtin, A. and Ashcroft,N.W. (1985), {\it Phys.Rev.A} {\bf 32}, 
2909-2919.
\item{} Deblieck, R. and Lekkerkerker, H.N.W. (1980), {\it J. de Physique Lett.}
{\bf 41}, L351-L355.
\item{} De Gennes, P.G. (1977), {\it J. de Physique  Lett.} {\bf 38}, L441-L443.
\item{} De Gennes, P.G. (1979), {\it Scaling Concept in Polymer  Physics},
Cornell University Press.
\item{} De Gennes, P.G. and Prost, J. (1993), {\it  The Physics of Liquid
Crystals}, Clarendon Press- Oxford.
\item{} Denton,A.R. and Ashcroft, N.W. (1989), {\it Phys.Rev. A} 
{\bf 39}, 4701-4708.
\item{} Denton,A.R. and Ashcroft, N.W. (1990), {\it Phys.Rev. A} {\bf 42}, 7312-7329; 
\item{} Denton,A.R. and Ashcroft, N.W. (1991), {\it Phys.Rev. A} {\bf 44}, 8242-8248.
\item{}  Des Cloiseaux, J. and Jannik, G. (1990), {\it Polymers in solutions:
their modelling and structure},  Oxford - Clarendon Press.
\item{} Deutsch, H.-P. and Binder, K., (1992), {\it 
Macromolecules} {\bf 25},  6214-6230.
\item{} Dijkstra, M. and Frenkel, D. (1994), {\it Phys.Rev.Lett.} {\bf 72},
298-300.
\item{} Dijkstra, M., Frenkel, D. and Hansen,  J.P. (1994), {\it J.Chem.Phys},
{\bf 101}, 3179-3189.
\item{} Doane,  J.W. (1990), in {\it Liquid  Crystals: Applications and Uses},
ed. Bahadur, B. World Scientific 362-396.
\item{} Doi,M.  and Edwards, S.F. (1986), {\it The Theory of Polymer Dynamics},
Clarendon, Oxford.
\item{} Dorgan, J.R. and Soane, D.S. (1990), {\it Mol.Cryst.Liq.Cryst.}
{\bf 188}, 129-146.
\item{} Edwards, S.F. (1966), {\it Proc.Phys.Soc.} {\bf 88}, 265-280.
\item{} Edwards, S.F. (1975), {\it J.Phys. A}, {\bf 8}, 1670-1680.
\item{} Evans, R. (1979), {\it Adv.Phys.} {\bf 28}, 143-200.
\item{} Evans, R. and Marconi, M.B. (1987), {\it J.Chem.Phys.}
{\bf 86}, 7138-7148.
\item{} Flory, P. (1953), {\it Principles of Polymer Chemistry},
Cornell University Press, Ithaca, NY.
\item{} Floudas, G.,  Pakula, T., Stamm, M. and Fischer, E.W.,
(1993),\hfill\break {\it Macromolecules}, {\bf 26}, 1671-1675.
\item{} Frenkel, D. (1991), in {\it Liquids, Freezing and Glass Transition}
ed by Hansen, J.P., Levesque, D., and Zinn-Justin, J. (Les Houches session LI)
North Holland, 689-762. 
\item{} Frenkel, D. (1994), {\it J.Phys.Condens. Matter} {\bf 6}, A71-A78.
\item{} Frenkel, D. and Louis, A.A. (1992) {\it Phys.Rev.Lett} {\bf 68}, 3363-3365.
\item{} Furman, D., Dattagupta, S., and Griffiths, R.B.(1977),
{\it Phys.Rev. B} {\bf 15}, 441-464. 
\item{} Gehlsen, M., Rosendale, J.H., Bates,F.S.,
Wignall, G.D., Lotte, H. and  Almdal, K., 
(1992) {\it Phys.Rev.Lett.} {\bf 68}, 2452-2455.
\item{} Gompper, G. and Schick, M. (1994), {\it Self-Assembling Amphiphilic Systems},
vol 16 of {\it Phase Transitions and Critical Phenomena} ed. Domb, C. and Lebowitz, J.L.
Academic Press, 1-181.
\item{} Grant,D.J.W., Higuchi, T.,  (1990), {\it Solubility Behavior of
Organic Compounds}, John Wiley \& Sons Inc.
\item{} Hagen, M.H.J., Meijer, E.J., Mooij, G.C.A.M., Frenkel, D., 
Lekkerkerker, H.N.W., (1993), {\it Nature}, {\bf 365}, 425-426.
\item{} Hansen, J.P. and  McDonald, I.R. (1986), {\it Theory of Simple
Liquids}, Academic Press.
\item{} Henderson, D. and Leonard, P.J. (1971) in {\it Physical Chemistry
(an advanced treatise)} Vol. VIIIB ed. Henderson, D., 
Academic Press, 414-510.
\item{} Hildebrand, J.H., Scott, R.L. (1950), 
{\it Solubility of Nonelectrolytes} 3rd ed., Reinhold Publishing Corp.
\item{} Hobbie, E.K., Bauer, B.J.  and Han, C.C. (1994), {\it Phys.Rev.Lett},
{\bf 72}, 1830-1833.
\item{} Ho\l yst, R. and Vilgis, T.A.  
(1994), {\it Phys.Rev.E} {\bf 50},  2087-2092.
\item{} Ho\l yst, R. and Vilgis, T.A. 
(1993), {\it J.Chem.Phys.} {\bf 99}, 4835-4844.
\item{} Ho\l yst, R. and Schick, M. (1992), {\it J.Chem.Phys.}, {\bf 96}, 721-729.
\item{} Ho\l  yst, R. and Schick, M. (1992), {\it J.Chem.Phys.}, {\bf 96}, 7728-7737.
\item{} Israelachvili, J.N., (1985), {\it Intermolecular Interactions
And Surface Forces}, Academic Press.
\item{} James, K.C. (1986), {\it Solubility and related properties},
Dekker, New York.
\item{} Jones, R.A.L., Norton, L.J., Kramer, E.J., Bates, F.S., Wiltzius, P.
(1991), {\it Phys.Rev.Lett}, {\bf 66}, 1326-1329.
\item{} Koga, T. and Kawasaki, K. (1993), {\it Physica A}, {\bf 196}, 389-415.
\item{} Landau, L.D.  and Lifshitz, E.M. (1980), {\it Statistical Physics}
3rd Edition part 1, Pergamon Press.
\item{} Langer, J.S. (1992)
in {\it Solids far from Equilibrium} ed. C.Godr\`eche, Cambridge University
Press, 298-362.
\item{} Laughlin, R.G. (1994), {\it The Aqueous Phase Behavior of Surfactants},
Academic Press.
\item{} Leibler, L. (1980) {\it Macromolecules}, {\bf 13}, 1602-1617.
\item{} Lifschitz, M., Dudowicz, J. and Freed, K.F. (1994), {\it J.Chem.Phys.}
3957-3978.
\item{} L\"owen, H. (1994) {\it Phys.Rep.} {\bf 237}, 251-324.
\item{} Ma\"issa, P. and Sixou, P. (1989), {\it Liquid Crystals}, {\bf 5}, 1861-1873.
\item{} Matsen, M.W. and Schick, M. (1994), {\it Phys.Rev.Lett.}, {\bf 72}, 2660-2663.
\item{} Mc-Graw Hill Dictionary of Chemical Terms (1984) 3rd ed. 
\item{} Meijer, E.J. and Frenkel, D. (1994), {\it J.Chem.Phys.} {\bf 100},
6873-6887.
\item{} Nernst, W. (1904) {\it  Theoretical Chemistry}, MacMillan and Co.,Ltd.
\item{} Onsager, L. (1949) {\it Proc. NY Acad. Sci.}, {\bf 51}, 627-655.
\item{} Poniewierski, A. and Ho\l yst, R. (1988), {\it Phys.Rev.Lett. },
{\bf 61}, 2461-2464.
\item{} Poniewierski, A. and Ho\l yst, R. (1990), {\it Phys.Rev. A} {\bf 41},
6871-6880.
\item{} Proceedings of the Int. School. of Physics, (1985) Course XC {\it
Physics of Amphiphiles: Micelles, Vesicles and Microemulsions} ed.
Degiorgio, V., North Holland.
\item{} Roe,R-J. and Zin, W-C.  (1984), {\it Macromolecules} {\bf 17}, 189-194.
\item{} Rowlinson, J.S. and Swinton, F.L. (1982),
{\it Liquids and Liquid Mixtures}, Butterworth Scientific.
\item{} Schwahm, D., Mortensen, K. and Yee-Madeira, H. (1987), {\it Phys.Rev.Lett },
{\bf 58}, 1544-1546.
\item{} Schweizer, K.S. and Curro, J.G. (1988), {\it Phys.Rev.Lett.} {\bf 60}, 809-812.
\item{} Shaw, M.R. and Thirumalai, D. (1991), {\it  Phys.Rev. A}, {\bf 44}, R4797-R4800.
\item{} Sigaud, G., Nguyen, H.T., Achard, M.F. and
Twieg, R.J. (1990), {\it Phys.Rev.Lett}, {\bf 65}, 2796-2799.
\item{} Sivardi\'ere, J. (1980), {\it J. de Physique} {\bf 41},  1081-1089.
\item{} Tanaka, H. (1994), {\it Phys.Rev.Lett}, {\bf 72}, 1702-1705.
\item{}  Tanaka, H. (1993), {\it Phys.Rev.Lett.}, {\bf 70}, 2770-2773, 3524.
\item{} Tarazona, P. (1985)  {\it Phys.Rev.A}  {\bf 31}, 2672-2679.
\item{} Tejero, C.F., Daanoum, A., Lekkerkerker, H.N.W. and Baus, M.
(1994)\hfill\break {\it Phys.Rev.Lett.} {\bf 73}, 752-755.
\item{} Teubner, M and Strey, R. (1987), {\it J.Chem.Phys.}, {\bf 87}, 
3195-3200.
\item{} Van Duijneveldt, J.S. and Lekkerkerker, H.N.W. (1993), 
{\it Phys.Rev.Lett.} {\bf 71}, 4264-4266.
\item{} Van Oss, C.J., Absolom, D.R. and Neumann, A.W. (1980),
{\it Colloids Surf}. {\bf 1}, 45-56.
\item{} Van Roij, R. and Mulder, B. (1994), {\it J.Phys. II France}, {\bf 4}, 1763-1769.
\item{} Walker,  J.S. and Vause, C.A. (1983), {\it J.Chem.Phys.} {\bf 79}, 
2660-2676.
\item{} Wiltzius, P.  and Cumming, A. (1991), {\it Phys.Rev.Lett}, {\bf 66},
3000-3003.
\item{} Xu,H. and Baus, M. (1987), {\it J.Phys.C} {\bf 20}, L373-L380.
\item{} Xu,H. and Baus, M. (1992), {\it J.Phys.Condens.Matter} {\bf 4}, L663-L668.

\noindent{\bf Figure Caption}

\item{Fig.1} The pressure composition phase diagram for the ideal binary 
mixture (A,B components) (see Eq.(5)). The region between the solid lines
is the two-phase region where vapor and liquid coexist. 
Along the upper curve evaporation of the liquid starts 
and along the lower curve the 
condensation of the vapor starts. 
\item{Fig.2} The temperature composition phase diagram for the 
binary mixture.
The dashed line shows how the process of fractional 
distillation proceeds. In this way we can obtain almost pure (B rich)
liquid.
\item{Fig.3} A phase diagram with a high boiling 
or negative azeotrope (maximum on the temperature-composition diagram). 
At this point liquid boils without changing its composition.
This type of diagram occurs for example for water/nitric acid,
chloroforme/acetone and hydrochloric acid/water.
\item{Fig.4} A phase diagram with a low-boiling 
or positive 
azeotrope (minimum on the temperature-composition diagram). 
This type of diagram can occur
for water/ethanol, dioxane/water. 
\item{Fig.5} Typical phase diagram for the
partially miscible liquids with the upper critical point (maximum on the
coexistence curve) (e.g. methanol and hydrocarbon liquids).
\item{Fig.6} Typical phase diagram for the partially miscible
liquids with a lower critical point (minimum on the coexistence curve)
(e.g. water and diethylamine or triethylamine).
\item{Fig.7} The phase diagram with the upper and lower critical points
(e.g. nicotine and water ).
\item{Fig.8} The phase diagram with a lower and upper critical point
and the miscibility gap between these point (e.g. acetone and polystyrene).
\item{Fig.9} The phase diagram for the mixture, where the upper 
critical point is below the
boiling curve.
\item{Fig.10} The phase diagram for the mixture, not  
fully miscible (i.e. in all proportions) in the
liquid state (e.g. water/diethyl ether, water/chloroform 
and high hydrocarbons and methanol).
\item{Fig.11} The clathrate cage around the non hydrogen bonding solute molecule.
The lines indicate the hydrogen bonds between the solvent molecules.
\item{Fig.12} Lamellar phase 
(periodic in one direction) of the AB diblock copolymer. The dots
denote the position of the points where A block (dashed line)
is chemically joined 
to the B  block (solid line).
\item{Fig.13} The cubic 
bicontinuous phase in the amphiphilic system. Symmetry Pm3m. 
Known as the Schwartz P triply periodic minimal surface.
Here  only the unit cell is shown.
The internal interfaces are  shown only. In the binary mixture
of water and amphiphiles, water is on both sides of the surface.
In the ternary mixture of oil, water and amphiphile
we have
oil on one side of the  surface and water on the other. 
The interface is formed by the amphiphilic molecules.
Water (oil) form interconnected channels
which span the whole volume. 
\item{Fig.14} The cubic bicontinuous phase in the amphiphilic system.
Symmetry F\=43m. Known as the Schwartz D triply periodic minimal surface.
Legend as in Fig.13.
\item{Fig.15} The gyroid structure., Known as the Schoen G surface.
Symmetry Ia3d. Legend as in Fig.13.
\vfill\eject\end